\documentclass[11pt]{article}

\title{Learning Robust Hybrid Control Barrier Functions\\ for Uncertain  Systems}

\usepackage{amsmath}
\usepackage{amssymb}
\usepackage{mathtools}
\usepackage{tikz}
\usepackage{pgfplots}
\usepackage{graphicx}
\usepackage{epstopdf}
\usepackage{setspace}
\usepackage[caption=false,font=footnotesize]{subfig}
\usepackage{xcolor}
\allowdisplaybreaks  
\usepackage{cite}
\usepackage{algorithmicx}
\usepackage{algpseudocode}
\usepackage{algorithm}
\usepackage{bbm}
\usepackage{makecell}
\usepackage{authblk}
\usepackage{fullpage,etoolbox}
\usepackage{hyperref}

\allowdisplaybreaks

\usepackage{amsthm}

\newtheorem{definition}{Definition} 
\newtheorem{theorem}{Theorem} 
\newtheorem{proposition}{Proposition} 
\newtheorem{assumption}{Assumption} 
\newtheorem{remark}{Remark}
\newtheorem{lemma}{Lemma}

\usepackage{enumitem}
\setlist{nosep}
\author[1]{Alexander Robey\thanks{Alexander Robey and Lars Lindemann contributed equally.}}
\author[1]{Lars Lindemann$^{\ast}$}
\author[2]{Stephen Tu}
\author[1]{Nikolai Matni}
\affil[1]{Department of Electrical and Systems Engineering, University of Pennsylvania}
\affil[2]{Google Brain Robotics}

\date{\today}

\begin{document}



\maketitle

\begin{abstract}
The need for robust control laws is especially important in safety-critical applications. We propose robust hybrid control barrier functions as a means to synthesize control laws that ensure robust safety. Based on this notion, we formulate an optimization problem for learning robust hybrid control barrier functions from data. We identify sufficient conditions on the data such that feasibility of the optimization problem ensures correctness of the learned robust hybrid control barrier functions. Our techniques allow us to safely expand the region of attraction of a compass gait walker that is subject to model uncertainty.
\end{abstract}

\section{Introduction}
\label{sec:introduction}

{Robust control} explicitly accounts for differences between the system and the {model of the system} that is used to design the control law. Such differences are referred to as {model uncertainties} and are inevitable due to modeling errors and the desire to keep the model complexity reasonable. Accounting for model uncertainties in the control design is hence pivotal, especially in safety-critical applications where we are interested in {robust safety}. Oftentimes, such safety-critical systems are hybrid, i.e., states may change continuously (``flow'') and discontinuously (``jump'') as time progresses. Examples include autonomous vehicles in urban areas \cite{schwarting2018planning}  and robots navigating a warehouse using semantic logic \cite{kress2009temporal}. Importantly, these systems share the similarities 
that: 1) {data} exhibiting safe behavior is readily available or easily collected, 2) in most cases, their hybrid system dynamics are well understood and can be modeled, and 3) {uncertainty sets} can be quantified expressing the confidence in the model. In this paper, we propose a data-driven and optimization-based approach to learn {safe and robust control laws} for uncertain hybrid systems using {robust hybrid control barrier functions} (RHCBF). 

\textbf{Related Work:} Safety can be defined as the ability of a system to avoid a set of {unsafe states}, e.g., states that violate a minimum safety distance.  Control barrier functions (CBFs) for continuous-time systems have been introduced in \cite{wieland2007constructive} and \cite{ames2017control} to render a set of {safe states} controlled forward invariant. A CBF allows us to define a set of safe control inputs, i.e., control inputs that keep the system within the set of safe states. CBFs have further been proposed for discrete-time systems \cite{agrawal2017discrete,ohnishi2019barrier} as well as for hybrid systems \cite{lindemann2020learning}. Barrier functions for hybrid systems, as a means to certifying safety,  have been introduced in \cite{maghenem2019characterizations,bisoffi2018hybrid}. Robust safety by means of CBFs has been considered in two directions. The works in \cite{kolathaya2018input} and \cite{xu2015robustness} consider a notion of input-to-state safety to quantify the safety violation in terms of the size of the model uncertainty when using CBFs. Importantly, safety is not guaranteed here. Conversely, \cite{jankovic2018robust} proposes robust CBFs for continuous-time systems to guarantee robust safety by accounting for all admissible model uncertainties. While such an approach is in general conservative, the author allows the use of an estimator of the model uncertainty, similar to our work, reducing conservatism.

While CBFs provide a theoretical formalism to ensure safety, the bottleneck is the construction of CBFs for general systems. The construction of polynomial barrier functions using sum-of-squares programming was proposed in \cite{prajna2007framework}. Finding CBFs poses additional challenges in terms of the control input resulting in bilinear sum-of-squares programming approaches \cite{ames2019control,xu2017correctness,wang2018permissive}. Such approaches only apply to polynomial systems and are subject to scalability issues. Recent effort has been made towards {learning CBFs from data}. In~\cite{yaghoubi2020training}, a deep neural network controller is trained to imitate a control law based on an existing CBF. In~\cite{srinivasan2020synthesis}, a CBF is synthesized from safe and unsafe data using support vector machines, while \cite{saveriano2019learning} cluster data and learn a linear CBF for each cluster. All of the aforementioned works present empirical validations of their methods while no formal correctness guarantees are provided. In~\cite{jin2020neural}, a Lyapunov, barrier, and
a policy function is learned from data and the correctness is verified post-hoc using Lipschitz arguments. In~\cite{ferlez2020shieldnn}, a method is proposed that learns a provably correct neural network safety guard for kinematic bicycle models. Also related is the work by \cite{khansari2014learning} in which motion primitives are learned from expert demonstrations that are stabilized by using a learned control Lyapunov functions. The authors in \cite{chen2020learning} propose a counter-example guided approach to learn Lyapunov functions for known closed-loop systems, while \cite{boffi2020learning} learn Lyapunov functions from data and without system knowledge. In our previous works \cite{robey2020learning,lindemann2020learning}, we propose a data-driven approach for learning CBFs for nonlinear and hybrid systems, respectively, assuming system knowledge. Alongside, we provide sufficient conditions ensuring correctness of the learned CBF using Lipschitz continuity and covering number arguments. However, all of the previous works on learning CBFs have not addressed robustness issues.

\textbf{Contributions:} We learn provably correct robust control barrier functions for uncertain hybrid systems from data. First, we define robust hybrid control barrier functions based on a flow and a jump constraint to enforce robust safety. We then formulate an optimization problem that incorporates these flow and jump constraints evaluated on safe data-points along with 1) constraints that shape the level sets of the RHCBF and 2) Lipschitz and boundedness constraints of the RHCBF. We provide sufficient conditions on the data under which a feasible solution to the optimization problem constitutes a valid RHCBF. To solve the optimization problem, we propose an unconstrained relaxation inspired by recent results on probably approximately correct constrained learning \cite{chamon2020probably}. Lastly, we present simulations on a compass gait walker that is subject to model uncertainty.


\section{Background and Problem Formulation}
\label{sec:backgound}


\emph{Notation: }Let $\text{dom}(z):=\{(t,j)\in \mathbb{R}_{\ge 0}\times\mathbb{N}| \exists \zeta\in\mathbb{R}^{n_z} \text{ s.t. } z(t,j)=\zeta\}$ be the domain of a function $z:\mathbb{R}_{\ge 0}\times\mathbb{N}\to  \mathbb{R}^{n_z}$. A continuous function $\alpha:\mathbb{R}\to\mathbb{R}$ is an extended class $\mathcal{K}$ function if $\alpha$ is strictly increasing and $\alpha(0)=0$. Let $\|\cdot\|$ be a norm and let $\|\cdot\|_\star$ denote its dual norm. For $\epsilon>0$, let $\mathcal{B}_{\epsilon}(z^i):=\{z\in\mathbb{R}^{n_z}\, \big{|}\, \|z-z^i\|\le \epsilon\}$ be a closed norm ball around $z^i\in\mathbb{R}^{n_z}$. Let bd$(\mathcal{C})$ and int$(\mathcal{C})$ be the boundary and interior of a set $\mathcal{C}$.  We denote a vector consisting of all ones by $\boldsymbol{1}$.

\subsection{Control Barrier Functions}
\label{sec:cbf}
At time $t\in\mathbb{R}_{\ge 0}$, let $x(t)\in\mathbb{R}^n$ be the state of the  system
\begin{align}\label{eq:system}
\dot{x}(t)=f(x(t))+g(x(t))u(x(t)), \; x(0)\in\mathbb{R}^n
\end{align}
where $f:\mathbb{R}^n\to\mathbb{R}^n$ and $g:\mathbb{R}^n\to\mathbb{R}^{n\times m}$ are continuous functions. Let solutions to \eqref{eq:system} under a continuous control law $u:\mathbb{R}^n\to\mathbb{R}^m$ be $x:\mathcal{I}\to \mathbb{R}^n$ where $\mathcal{I}\subseteq\mathbb{R}_{\ge 0}$ is the maximum definition interval of $x$. Consider  a continuously differentiable function $h:\mathbb{R}^n\to\mathbb{R}$ and define the set 
\begin{align*}
    \mathcal{C}:=\{x\in\mathbb{R}^n \, \big{|} \, h(x)\ge 0\}
\end{align*} 
that we aim to render forward invariant for the system in \eqref{eq:system} through an appropriate choice of control law $u$. Note that $\mathcal{C}$ is closed and further assume that $\mathcal{C}$ is not the empty set. Now, let $\mathcal{D}$ be an open set that is such that $\mathcal{D}\supseteq \mathcal{C}$. The function $h(x)$ is said to be a \emph{control barrier function} on $\mathcal{D}$ if there exists a locally Lipschitz continuous extended class $\mathcal{K}$ function $\alpha:\mathbb{R}\to\mathbb{R}$ such that 
\begin{align*}
\sup_{u\in \mathcal{U}} 
\langle \nabla h(x), f(x)+g(x)u\rangle \ge -\alpha(h(x))
\end{align*} 
holds for all $x\in\mathcal{D}$, where $\mathcal{U}\in\mathbb{R}^m$ defines constraints on the control input $u$. We define the set of \emph{safe control inputs} induced by a CBF $h(x)$ to be 
$
K_{\text{CBF}}(x):=\{u\in\mathbb{R}^m \, \big{|} \, 
\langle \nabla h(x), f(x)+g(x)u\rangle \ge -\alpha(h(x))\}$. The next result follows in the spirit of \cite{ames2017control} and is provided in \cite{lindemann2020learning} without requiring the regularity assumption that $\nabla h(x)\neq 0$ when $x\in \text{bd}(\mathcal{C})$.
\begin{lemma}\label{lem:1}
Assume that $h(x)$ is a  control barrier function on $\mathcal{D}$ and that $u:\mathcal{D}\to \mathcal{U}$ is a continuous function with $u(x)\in K_{\text{CBF}}(x)$. Then $x(0)\in\mathcal{C}$ implies $x(t)\in\mathcal{C}$ for all $t\in \mathcal{I}$. If $\mathcal{C}$ is compact, it follows that $\mathcal{C}$ is forward invariant under $u(x)$, i.e., $\mathcal{I}=[0,\infty)$.
\end{lemma}

\subsection{Hybrid Systems }
\label{sec:RRSTL}

We model and analyze hybrid systems using the formalism of~\cite{goebel2012hybrid}.

\begin{definition}\label{def:hybrid_system}
	A hybrid system \cite{goebel2012hybrid} is a tuple $\mathcal{H}:=(C,F,D,G)$ where $C\subseteq\mathbb{R}^{n_z}$, $D\subseteq\mathbb{R}^{n_z}$, $F: \mathbb{R}^{n_z}\times \mathbb{R}_{\ge 0}\times\mathbb{N}\to \mathbb{R}^{n_z}$, and $G:\mathbb{R}^{n_z}\times \mathbb{R}_{\ge 0}\times\mathbb{N}\to \mathbb{R}^{n_z}$ are the flow and jump sets and the continuous flow and jump maps, respectively. At the hybrid time $(t,j)\in \mathbb{R}_{\ge 0}\times\mathbb{N}$, let $z(t,j)\in \mathbb{R}^{n_z}$ be the hybrid state with initial condition $z(0,0)\in C\cup D$ and the hybrid system dynamics
	\begin{subequations}\label{eq:hybrid_systems}
	\begin{align}
	\dot{z}(t,j)&= F(z(t,j),t,j) \ \text{  for  } \ z(t,j) \in C,\\ 
	 z(t,j+1)&= G(z(t,j),t,j) \ \text{  for  }\ z(t,j)\in D.
	\end{align}
	\end{subequations}
\end{definition}
Note that the above definition is, without much change, a time-varying version of the hybrid systems formalism presented in \cite{goebel2012hybrid}. Solutions to \eqref{eq:hybrid_systems} are parameterized by $(t,j)$, where $t$ indicates continuous \emph{flow} according to $F(z,t,j)$ and $j$ indicates discontinuous \emph{jumps} according to $G(z,t,j)$. Now let $\mathcal{E} \subseteq \mathbb{R}_{\ge 0}\times \mathbb{N}$ be a \emph{hybrid time domain} \cite[Ch. 2.2]{goebel2012hybrid}, i.e., $\mathcal{E}$ is an infinite union of intervals of the form $[t_j,t_{j+1}]\times \{j\}$ or a finite union of intervals of the form $[t_j,t_{j+1}]\times \{j\}$ where the last interval, if it exists, has the form $[t_j,t_{j+1}]\times \{j\}$, $[t_j,t_{j+1})\times \{j\}$, or $[t_j,\infty)\times \{j\}$. 

\begin{definition}\label{def:hybrid_solutions}
	A function $z:\mathcal{E}\to  C \cup D$ is a hybrid solution to $\mathcal{H}$ if $z(0,0)\in C \cup D$ and 
	\begin{itemize}[leftmargin=*]
	    \item for each $j\in\mathbb{N}$ such that $I_j:= \{t\in\mathbb{R}_{\ge 0}|(t,j)\in \text{dom}(z) \}$ is not a singleton, $z(t,j)\in C$ and  $\dot{z}(t,j)=F(z(t,j),t,j)$ for all $t\in[\min I_j, \sup I_j)$
	    \item for each $(t,j)\in\text{dom}(z)$ s.t. $(t,j+1)\in\text{dom}(z)$, $z(t,j)\in D$ and $z(t,j+1)=G(z(t,j),t,j)$.
	\end{itemize}
\end{definition}


\subsection{Problem Formulation}
The class of \emph{hybrid control systems} that we consider is
\begin{subequations}\label{eq:hybrid_control_systems}
\begin{align}
&\begin{rcases}  
\dot{z}(t,j)&= f_c(t,j)+g_c(t,j)u_c(t,j)   \\
f_c(t,j)&:=F_c(z(t,j),t) \hspace{0.2cm}\text{(internal dynamics)}\\
g_c(t,j)&:=G_c(z(t,j),t) \hspace{0.5cm}\text{(input dynamics)}\\
u_c(t,j)&:=U_c(z(t,j),t) \hspace{1.21cm}\text{(control law)}
\end{rcases} \text{ for }z(t,j)\in C\\
&\begin{rcases}
z(t,j+1)&\hspace{-0.35cm}= f_d(t,j)+g_d(t,j)u_d(t,j) \\
f_d(t,j)&\hspace{-0.35cm}:=F_d(z(t,j),t) \hspace{0.2cm}\text{(internal dynamics)}\\
g_d(t,j)&\hspace{-0.35cm}:=G_d(z(t,j),t) \hspace{0.5cm}\text{(input dynamics)}\\
u_d(t,j)&\hspace{-0.35cm}:=U_d(z(t,j),t) \hspace{1.21cm}\text{(control law)}
\end{rcases}\text{ for }z(t,j)\in D
\end{align}
\end{subequations}
where the functions $U_c:C\times \mathbb{R}_{\ge 0} \to \mathcal U_c \subseteq \mathbb{R}^{m_c}$ and $U_d:D\times \mathbb{R}_{\ge 0} \to \mathcal U_d \subseteq\mathbb{R}^{m_d}$ are continuous in the first and piecewise continuous and bounded in the second argument. The functions $U_c$ and $U_d$ define the control laws, while the sets $\mathcal{U}_c$ and $\mathcal{U}_d$ impose input constraints. The functions $F_c:\mathbb{R}^{n_z}\times \mathbb{R}_{\ge 0}\to\mathbb{R}^{n_z}$, $F_d:\mathbb{R}^{n_z}\times \mathbb{R}_{\ge 0}\to\mathbb{R}^{n_z}$, $G_c:\mathbb{R}^{n_z}\times \mathbb{R}_{\ge 0}\to\mathbb{R}^{n_z\times m_c}$, and $G_d:\mathbb{R}^{n_z}\times \mathbb{R}_{\ge 0}\to\mathbb{R}^{n_z\times m_d}$ are in general \emph{only partially known} and  locally Lipschitz continuous in the first and piecewise continuous and bounded in the second argument. Define the combined internal and input dynamics by the functions $W_c:C\times\mathbb{R}_{\ge 0}\times \mathcal{U}_c\to\mathbb{R}^{n_z}$ and $W_d:D\times\mathbb{R}_{\ge 0}\times \mathcal{U}_d\to\mathbb{R}^{n_z}$ with
\begin{align*}
    W_c(z,t,u_c)&:=F_c(z,t)+G_c(z,t)u_c,\\
    W_d(z,t,u_d)&:=F_d(z,t)+G_d(z,t)u_d.
\end{align*}
These functions are  again  only partially known due to, for instance, unmodeled internal or input dynamics, modeling errors, or noise affecting the internal or input dynamics. 
\begin{assumption}\label{ass:1}
We assume to have estimates $\hat{W}_c(z,t,u_c)$ of $W_c(z,t,u_d)$ and $\hat{W}_d(z,t,u_c)$ of $W_d(z,t,u_d)$ together with functions $\Delta_c:C\times\mathbb{R}_{\ge 0}\times\mathcal{U}_c\to\mathbb{R}_{\ge 0}$ and $\Delta_d:D\times\mathbb{R}_{\ge 0}\times\mathcal{U}_d\to\mathbb{R}_{\ge 0}$ that bound the errors between $\hat{W}_c(z,t,u_c)$ and $W_c(z,t,u_c)$ as well as $\hat{W}_d(z,t,u_d)$ and $W_d(z,t,u_d)$ as
\begin{align*}
    \|\hat{W}_c(z,t,u_c)-W_c(z,t,u_c)\|&\le \Delta_c(z,t,u_c) \text{ for all } (z,t,u_c)\in C\times \mathbb{R}_{\ge 0}\times\mathcal{U}_c,\\
    \|\hat{W}_d(z,t,u_d)-W_d(z,t,u_d)\|&\le \Delta_d(z,t,u_d) \text{ for all }(z,t,u_d)\in D\times \mathbb{R}_{\ge 0}\times\mathcal{U}_d.
\end{align*} 
The functions $\hat{W}_c:C\times\mathbb{R}_{\ge 0}\times\mathcal{U}_c\to\mathbb{R}^{n_z}$, $\hat{W}_d:D\times\mathbb{R}_{\ge 0}\times \mathcal{U}_d\to\mathbb{R}^{n_z}$, $\Delta_c(z,t,u_c)$, and $\Delta_d(z,t,u_d)$ are assumed to be locally Lipschitz continuous in the first and piecewise continuous and bounded in the second argument.
\end{assumption}

Such estimates $\hat{W}_c(z,t,u_c)$ and $\hat{W}_d(z,t,u_d)$ may represent estimated internal and input dynamics of the system \eqref{eq:hybrid_control_systems}, e.g., by identifying model parameters,  together with a confidence estimate in the form of the error bounds $\Delta_c(z,t,u_c)$ and $\Delta_d(z,t,u_d)$. Let us now define the sets of \emph{admissible system dynamics} according to Assumption \ref{ass:1}
\begin{align*}
    \mathcal{W}_c(z,t,u_c)&:=\{w\in\mathbb{R}^{n_z}|\|\hat{W}_c(z,t,u_c)-w\|\le \Delta_c(z,t,u_c)\}\\
    \mathcal{W}_d(z,t,u_d)&:=\{w\in\mathbb{R}^{n_z}|\|\hat{W}_d(z,t,u_d)-w\|\le \Delta_d(z,t,u_d)\}.
\end{align*}

We do not assume completeness of the system in \eqref{eq:hybrid_control_systems} under control laws $U_c(z,t)$ and $U_d(z,t)$  and  system dynamics $W_c(z,t,U_c(z,t))\in\mathcal{W}_c(z,t,U_c(z,t))$ and $W_d(z,t,U_d(z,t))\in\mathcal{W}_d(z,t,U_d(z,t))$. We will, however, enforce this property. Completeness means that the hybrid time domain $\text{dom}(z)$ is unbounded (see \cite[Ch. 2.2]{goebel2012hybrid} for a formal definition). 

\begin{remark}
Note that Assumption \ref{ass:1} admits a \emph{very general formulation of an unknown system}. A common subcase is obtained for additive unmodeled internal dynamics or unmodeled disturbances. Then the functions $G_c(z,t)$ and $G_d(z,t)$ are known while the functions $F_c(z,t)=F_c^{\text{k}}(z,t)+F_c^{\text{uk}}(z,t)$ and $F_d(z,t)=F_d^{\text{k}}(z,t)+F_d^{\text{uk}}(z,t)$ are only partially known (here k abbreviates `known' while uk abbreviates `unknown'). We then only need estimates $\hat{F}_c^\text{uk}(z,t)$ and $\hat{F}_d^\text{uk}(z,t)$ so that the error bounds  $\Delta_c(z,t)$ and $\Delta_d(z,t)$ are independent of the control inputs $u_c$ and $u_d$ and such that $\|\hat{F}_c^{\text{uk}}(z,t)-F_c^{\text{uk}}(z,t)\|\le \Delta_c(z,t)$ for all $(z,t)\in C\times\mathbb{R}_{\ge 0}$ and $\|\hat{F}_d^{\text{uk}}(z,t)-F_d^{\text{uk}}(z,t)\|\le \Delta_d(z,t)$ for all $(z,t)\in D\times\mathbb{R}_{\ge 0}$. We remark that Assumption \ref{ass:1} even holds without estimates $\hat{F}_c^{\text{uk}}(z,t)$ and $\hat{F}_d^{\text{uk}}(z,t)$, i.e., when $\hat{F}_c(z,t)=0$ and $\hat{F}_d(z,t)=0$. The bounds $\Delta_c(z,t)$ and $\Delta_d(z,t)$ then directly bound the admissible disturbances $F_c^{\text{uk}}(z,t)$ and $F_d^{\text{uk}}(z,t)$. Availability of such estimates will, however, greatly reduce conservatism.
\end{remark}

This paper is  concerned with the \emph{safety} of the system in \eqref{eq:hybrid_control_systems} by confining system trajectories $z(t,j)$ to the set $\mathcal{S}\subseteq \mathbb{R}^{n_z}$ which we refer to as the \emph{geometric safe set}, i.e., the set of safe states as naturally specified on a subset of
the system configuration space (e.g., to avoid collision, vehicles must maintain a minimum
separating distance). Let us now formally define what we mean by safety.
\begin{definition}
A set $\mathcal{C}\subseteq\mathbb{R}^{n_z}$ is said to be \emph{robustly controlled forward invariant} with respect to the system in \eqref{eq:hybrid_control_systems} if there exist feedback control laws $U_c(z,t)$ and $U_d(z,t)$ such that, for all initial positions $z(0,0)\in\mathcal{C}$ and for all admissible system dynamics $W_c(z,t,U_c(z,t))\in\mathcal{W}_c(z,t,U_c(z,t))$ and $W_d(z,t,U_d(z,t))\in\mathcal{W}_d(z,t,U_d(z,t))$, every solution $z(t,j)$ to \eqref{eq:hybrid_control_systems} under $U_c(z,t)$ and $U_d(z,t)$ is such that: 1) $z(t,j)\in\mathcal{C}$ for all $(t,j)\in \text{dom}(z)$, and 2) the hybrid time domain $\text{dom}(z)$ is unbounded. If the set $\mathcal{C}$ is additionally contained within the geometric safe set $\mathcal{S}$, i.e., $\mathcal{C}\subseteq\mathcal{S}$, we say that the system in \eqref{eq:hybrid_control_systems} is \emph{safe} under the \emph{safe control laws} $U_c(z,t)$ and $U_d(z,t)$.
\end{definition}

\begin{figure*}
\centering
\includegraphics[scale=0.25]{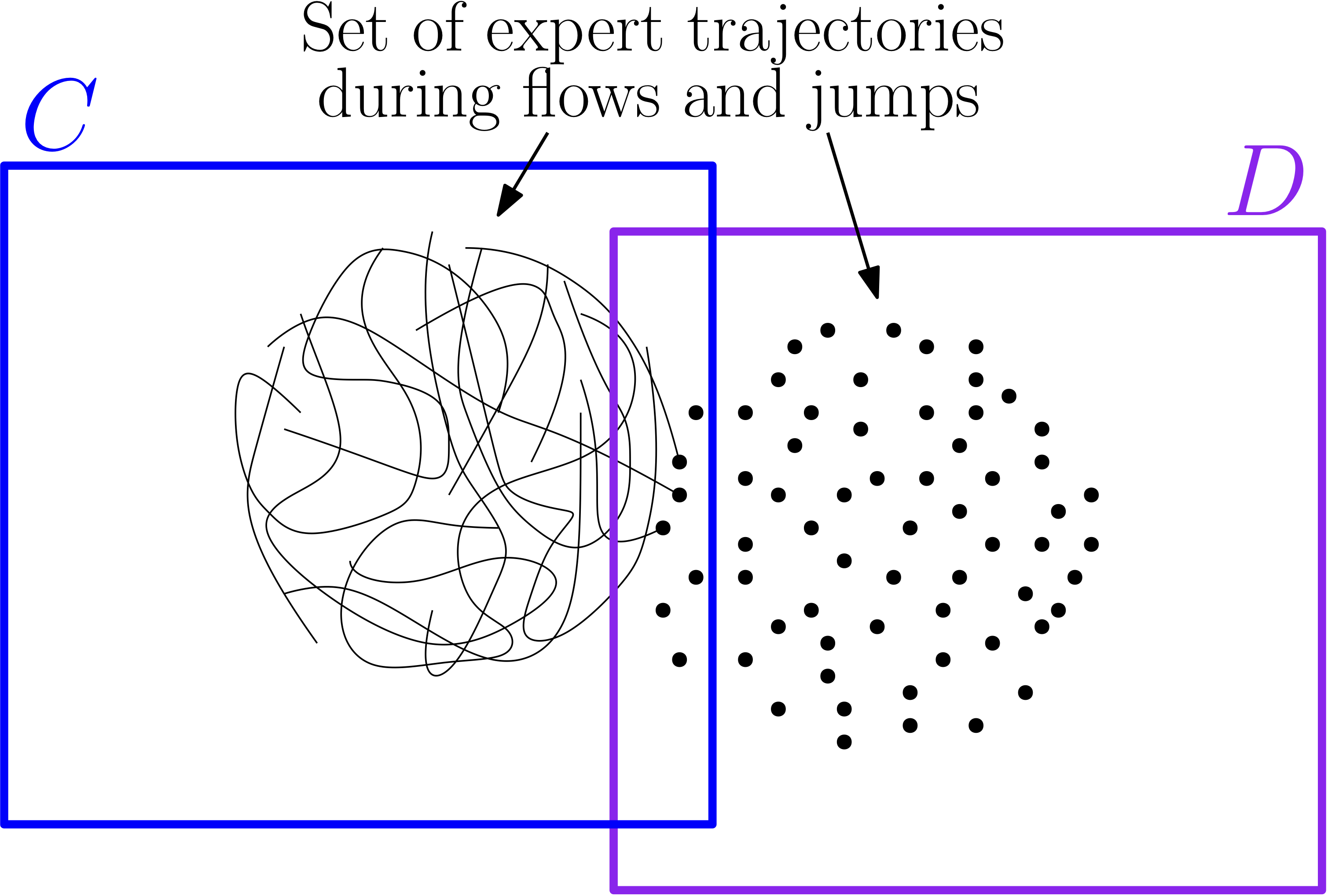}
\hspace{0.5cm}\includegraphics[scale=0.25]{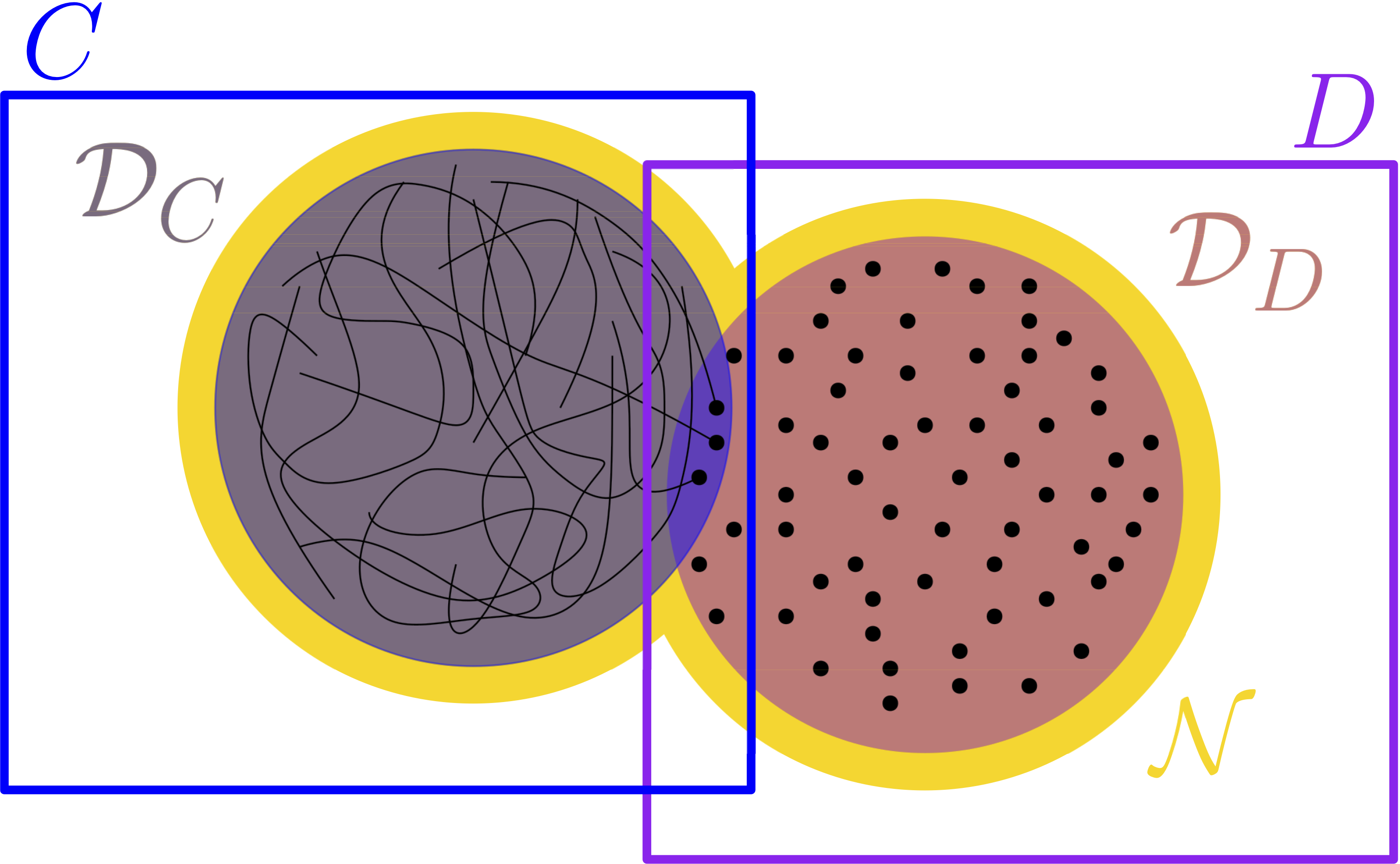}\hspace{0.5cm}\includegraphics[scale=0.25]{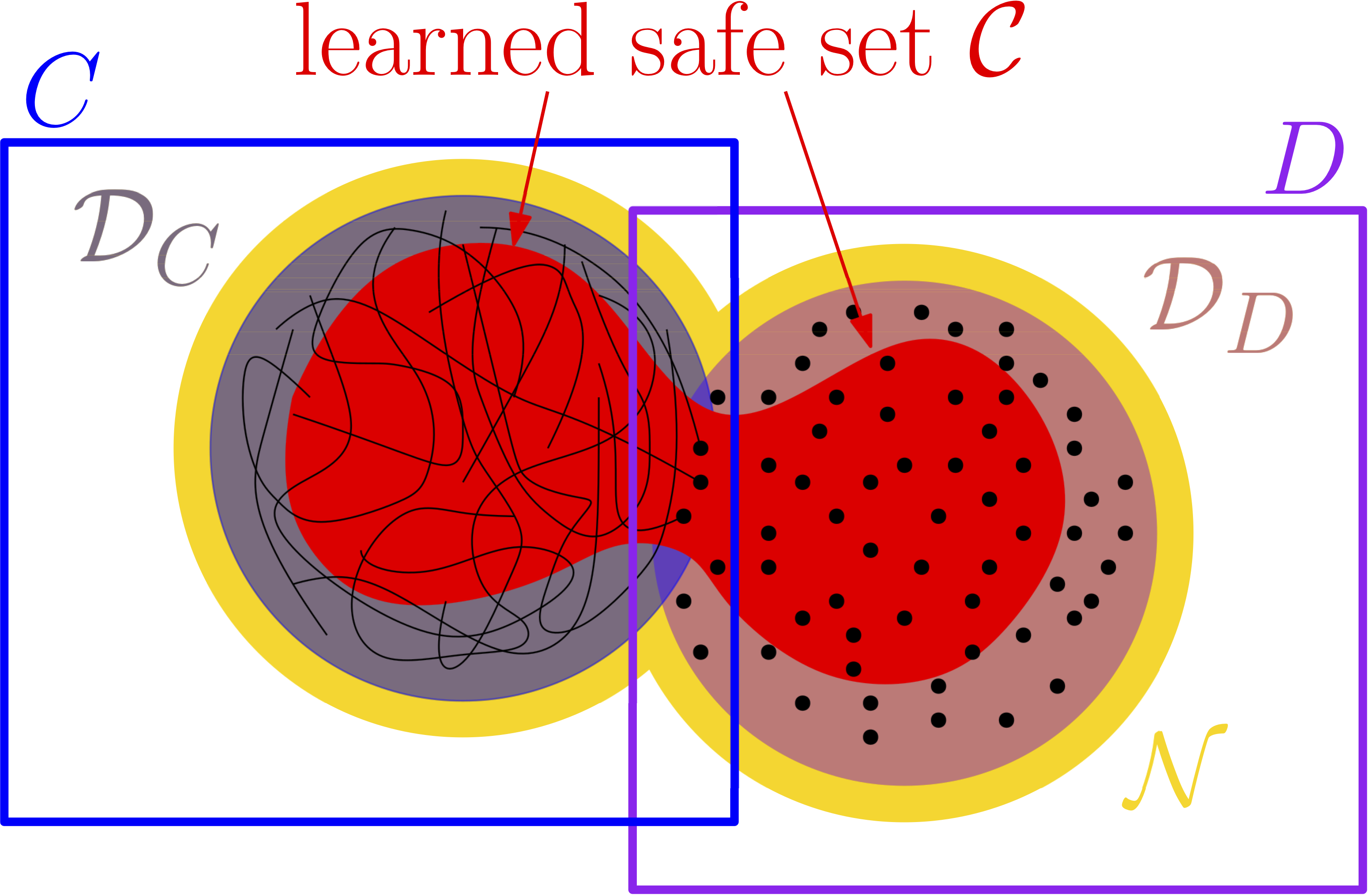}
\caption{Problem setup (left): The flow and jump sets $C$ and $D$ (blue and purple boxes) and the set of safe expert trajectories during flows and jumps (black lines and dots). Set definitions (middle): The sets $\mathcal{D}_C$ and $\mathcal{D}_D$ (black and light red balls) are the union of $\epsilon$ balls around the expert trajectories during flows and jumps. The set $\mathcal{N}$ (golden rings), defined around $\mathcal{D}_C$ and $\mathcal{D}_D$, ensures that the learned safe set $\mathcal{C}$ is such that $\mathcal{C}\subset\mathcal{D}\subseteq\mathcal{S}$. Note that the geometrical safe set $\mathcal{S}$ is not depicted here. Desired result (right): The learned safe set $\mathcal{C}$ (red region) is defined via the learned RHCBF $h(z)$.}
\label{fig:1}
\end{figure*}

Towards deriving safe control laws $U_c(z,t)$ and $U_d(z,t)$, we assume to be  given a set of \emph{expert trajectories} consisting of $N_c$ and $N_d$ discretely sampled data-points along flows and jumps as $Z_\text{dyn}^c:=\{(z^i,u^i_c,t^i)\}_{i=1}^{N_c}$ for $z^i\in C$, $u^i_c\in\mathcal{U}_c$ and $Z_\text{dyn}^d:=\{(z^i,u^i_d,t^i)\}_{i=1}^{N_d}$ for $z^i\in D$, $u^i_d\in\mathcal{U}_d$ as illustrated in Figure \ref{fig:1} (left). It is assumed that each $z^i\in\text{int}(\mathcal{S})$. A consequence of using expert trajectories is that each data-point in $Z_\text{dyn}^c$ and $Z_\text{dyn}^d$ corresponds to a specific realization of the admissible system dynamics.

Our goal is now to learn, from $Z_\text{dyn}^c$ and $Z_\text{dyn}^d$,  a twice continuously differentiable function $h:\mathbb{R}^{n_z}\to\mathbb{R}$ such that
\begin{align}\label{eq:set_C}
\mathcal{C}:=\{z\in\mathbb{R}^{n_z} \, \big{|} \, h(z)\ge 0\}
\end{align}
is a subset of the geometric safe set $\mathcal{S}$, i.e., $\mathcal{C}\subseteq\mathcal{S}$, 
and that $\mathcal{C}$ can be made robustly controlled forward invariant by appropriate safe control laws $U_c(z,t)$ and $U_d(z,t)$ that are defined implicitly via the function $h(z)$.

\section{Learning Robust Hybrid Control Barrier Functions (RHCBF) from Data}

We begin by defining the notion of a RHCBF as a means to synthesize safe control laws for the system \eqref{eq:hybrid_control_systems}.  We then show how such RHCBFs can be learned from data via a constrained optimization problem. Alongside, we provide sufficient conditions on the data $Z_\text{dyn}^c$ and $Z_\text{dyn}^d$ under which a feasible solution is a RHCBF.

\subsection{Robust Hybrid Control Barrier Functions (RHCBF)}
Let $h:\mathbb{R}^{n_z}\to\mathbb{R}$ be a twice continuously differentiable function for which the set $\mathcal{C}$ in \eqref{eq:set_C} is not empty and such that $\mathcal{C}\subseteq C\cup D$. The latter is a natural assumption since we are only interested in the domain $C\cup D$ where the system \eqref{eq:hybrid_control_systems} is defined. Let Lip$(h,z,\epsilon)$ denote the local Lipschitz constant within the set $\mathcal{B}_{\epsilon}(z)$, i.e., it holds that $|h(z')-h(z'')|\le $Lip$(h,z,\epsilon)\|z'-z''\|$ for all $z',z''\in \mathcal{B}_{\epsilon}(z)$. 

Consider now the sets $\mathcal{D}_C\subseteq C$ and $\mathcal{D}_D\subseteq D$ that are such that $\mathcal{C}\cap C \subseteq \mathcal{D}_C$ and $\mathcal{C}\cap D \subseteq\mathcal{D}_D$ from which it follows that $\mathcal{C}\subseteq \mathcal{D}_C\cup \mathcal{D}_D$, ensuring that the set $\mathcal{D}:=\mathcal{D}_C\cup\mathcal{D}_D$ fully covers $\mathcal{C}$ -- see Fig.~\ref{fig:1}(middle) and (right).  To avoid technicalities, assume  also that the set $\mathcal{D}_C$ is open.

\begin{definition}
The function $h(z)$ is said to be a \emph{robust hybrid control barrier function}  on $\mathcal{D}$  for the system \eqref{eq:hybrid_control_systems} if there exists a locally Lipschitz continuous extended class $\mathcal{K}$ function $\alpha:\mathbb{R}\to \mathbb{R}$ such that 
    \begin{align}
        \sup_{u_c\in \mathcal{U}_c} 
        &\langle \nabla h(z), \hat{W}_c(z,t,u_c)\rangle -\|\nabla h(z)\|_\star\Delta_c(z,t,u_c)  \ge -\alpha(h(z))  \text{ for all $(z,t)\in \mathcal{D}_C\times\mathbb{R}_{\ge 0}$,}\label{eq:cbf_const_hybrid}\\
        \sup_{u_d\in \mathcal{U}_d} &h(\hat{W}_d(z,t,u_d)) - \text{Lip}(h,\hat{W}_d(z,t,u_d),\Delta_d(z,t,u_d)) \Delta_d(z,t,u_d) \ge 0\text{ for all $(z,t)\in \mathcal{D}_D\times\mathbb{R}_{\ge 0}$.}\label{eq:cbf_disc_hybrid}
    \end{align}
\end{definition}
We define the sets of safe control inputs induced by a RHCBF $h(z)$ during flows and jumps to be
\begin{align*}
K_{\text{RHCBF},c}(z,t)&:=\{u_c\in\mathcal{U}_c\, \big{|} \,
\langle \nabla h(z), \hat{W}_c(z,t,u_c)\rangle  -\|\nabla h(z)\|_\star\Delta_c(z,t,u_c)\ge -\alpha(h(z))\},\\
K_{\text{RHCBF},d}(z,t)&:=\{u_d\in\mathcal{U}_d\, \big{|} \, 
 h(\hat{W}_d(z,t,u_d)) - \text{Lip}(h,\hat{W}_d(z,t,u_d),\Delta_d(z,t,u_d))\Delta_d(z,t,u_d)\ge 0\}.
 \end{align*}
\begin{theorem}\label{thm:1}
Assume that $h(z)$ is a robust hybrid control barrier function on $\mathcal{D}$ for the system \eqref{eq:hybrid_control_systems} and that $U_c:\mathcal{D}_C\times \mathbb{R}_{\ge 0}\to \mathcal{U}_c$ and $U_d:\mathcal{D}_D\times \mathbb{R}_{\ge 0} \to \mathcal{U}_d$ are continuous functions with $U_c(z,t)\in K_{\text{RHCBF},c}(z,t)$ and $U_d(z,t)\in K_{\text{RHCBF},d}(z,t)$. Then $z(0,0)\in\mathcal{C}$ implies $z(t,j)\in\mathcal{C}$ for all $(t,j)\in \text{dom}(z)$. If $\mathcal{C}$ is compact and satisfies $\mathcal{C}\subseteq C\cup D$,  the set $\mathcal{C}$ is robustly controlled forward invariant with respect to the system \eqref{eq:hybrid_control_systems} under control laws  $U_c(z,t)$ and $U_d(z,t)$.
\begin{proof}
First let us define the functions
\begin{align*}
\begin{rcases}
    {w}_c(t,j)&:={W}_c(z(t,j),t,u_c(t,j))\\
    {w}_d(t,j)&:={W}_d(z(t,j),t,u_d(t,j))
    \end{rcases} &\text{ true dynamics,}\\
\begin{rcases}
    \hat{w}_c(t,j)&:=\hat{W}_c(z(t,j),t,u_c(t,j))\\
    \hat{w}_d(t,j)&:=\hat{W}_d(z(t,j),t,u_d(t,j))
\end{rcases} &\text{ estimated dynamics,}\\
\begin{rcases}
    \delta_c(t,j)&:=\Delta_c(z(t,j),t,u_c(t,j))\\
    \delta_d(t,j)&:=\Delta_d(z(t,j),t,u_d(t,j)
\end{rcases} &\text{ error bounds.}
\end{align*}

During \emph{flows} with $I_j$ not being a singleton, and if $h(z(\min(I_j),j))\ge 0$, we can show that $h(z(t,j))\ge 0$ for all $t\in I_j$ as follows. Using the chain rule and since $u_c(t,j)\in K_{\text{RHCBF},c}(z(t,j),t)$ (due to $U_c(z,t)\in K_{\text{RHCBF},c}(z,t)$), note that the solution $z(t,j)$ is such that
  \begin{align}
 &\hspace{-0.5cm} \langle \nabla h(z(t,j)), \hat{w}_c(t,j)\rangle -\| \nabla h(z(t,j))\|_\star\delta_c(t,j)\ge-\alpha(h(z(t,j)))  \nonumber\\
\overset{(a)}{\Rightarrow} \;&\langle \nabla h(z(t,j)), w_c(t,j)\rangle \ge -\alpha(h(z(t))) \nonumber\\  
\Leftrightarrow \;&\dot{h}(z(t,j))\ge -\alpha(h(z(t,j))) \label{eq:thmmm}
\end{align}
for all $t\in [\min I_j, \sup I_j)$. The implication $(a)$ follows since 
\begin{align*}
   \| \nabla h(z(t,j))\|_\star\delta_c(t,j) \ge \| \nabla h(z(t,j))\|_\star\|\hat{w}_c(t,j)-w_c(t,j)\|
    \ge \langle\nabla h(z(t,j)),\hat{w}_c(t,j)-w_c(t,j)\rangle
\end{align*}
due to $w_c(t,j)\in\mathcal{W}_c(z(t,j),t,u_c(t,j))$. Next note that $\dot{v}(t)=-\alpha(v(t))$ with $v(0)\ge 0$ admits a unique solution $v(t)$ that is such that $v(t)\ge 0$ for all $t\ge 0$ \cite[Lemma 4.4]{Kha96}. Using the Comparison Lemma \cite[Lemma 3.4]{Kha96} and recalling that $h(z(\min(I_j),j))\ge 0$, it follows from \eqref{eq:thmmm} that $h(z(t,j))\ge v(t)\ge 0$ for all $t\in [\min I_j, \sup I_j)$. By continuity of $h(z)$ and $z(t,j)$ and since $\mathcal{C}$ is closed, it also holds that $h(z(\sup I_j,j))\ge 0$ if $I_j=[t_j,t_{j+1}]\times \{j\}$, i.e., the right end point is included in $I_j$, i.e., $z(0,0)\in\mathcal{C}$ implies $z(t,j)\in\mathcal{C}$ for all $t\in I_j$.

During \emph{jumps} and since $u_d(t,j)\in K_{\text{RHCBF},d}(z(t,j),t)$ (due to $U_d(z,t)\in K_{\text{RHCBF},d}(z,t)$), it holds that
\begin{align}\label{eq:disc_CBF}
\begin{split}
    &\hspace{-0.2cm}h(\hat{w}_d(t_{j+1},j))\ge\text{Lip}(h,\hat{w}_d(t_{j+1},j),\delta_d(t_{j+1},j))\delta_d(t_{j+1},j).
    \end{split}
\end{align}
Now adding $-h(w_d(t_{j+1},j))$ to both sides of the inequality in \eqref{eq:disc_CBF} and upper bounding the resulting left-hand side by 
\begin{align*}
&\text{Lip}(h,\hat{w}_d(t_{j+1},j),\delta_d(t_{j+1},j))\|\hat{w}_d(t_{j+1},j)-w_d(t_{j+1},j)\|\le \text{Lip}(h,\hat{w}_d(t_{j+1},j),\delta_d(t_{j+1},j))\delta_d(t_{j+1},j)
\end{align*}
by using Lipschitz continuity of $h(z)$ within the set $\mathcal{B}_{\delta_d(t_{j+1},j)}(\hat{w}_d(t_{j+1},j))$ and the fact that $w_d(t,j)\in\mathcal{W}_d(z(t,j),t,u_d(t,j))$,
leads to $h(z(t_{j+1},j+1))\ge 0$. Hence, $z(0,0)\in\mathcal{C}$ implies $z(t,j)\in\mathcal{C}$ for all $(t,j)\in \text{dom}(z)$.

We next need to show that the set $\mathcal{C}$ is robustly controlled forward invariant with respect to system \eqref{eq:hybrid_control_systems} if $\mathcal{C}$ is compact and if $\mathcal{C}\subseteq C\cup D$. Therefore, we  have to show that $\text{dom}(z)$ is unbounded. The main idea is to show that $z(t,j)$ can not leave the set $\mathcal{D}_C\cup\mathcal{D}_D$. The proof follows the same way as in \cite[Theorem 1]{lindemann2020learning} and is  omitted.
 \end{proof}
\end{theorem}

\subsection{Learning Robust Hybrid Control Barrier Functions}
\label{sec:compute}

Define the data sets $Z_\text{safe}^c := \{ z^i : (z^i, u^i_c,t^i) \in Z_\text{dyn}^c \}$ and $Z_\text{safe}^d := \{ z^i : (z^i, u^i_d,t^i) \in Z_\text{dyn}^d \}$ and, for $\epsilon_c,\epsilon_d>0$, let
\begin{align*}
    \mathcal{D}_D:=D\cap\bigcup_{z^i\in Z_\text{safe}^d}\mathcal{B}_{\epsilon_d}(z^i) \\ \mathcal{D}_C := \mathcal{D}_C' \backslash \mathrm{bd}(\mathcal{D}_C') \,\text{ where }\, \mathcal{D}_C'&:=C\cap \bigcup_{z^i\in Z_\text{safe}^c}\mathcal{B}_{\epsilon_c}(z^i)
\end{align*} 
be sets that need to be such that  $\mathcal{D}=\mathcal{D}_C\cup \mathcal{D}_D \subseteq \mathcal{S}$, which can be easily achieved even when data-points $z^i$ are close to $\text{bd}(\mathcal{S})$  by adjusting $\epsilon_c$ and $\epsilon_d$ or by omitting $z^i$. Note that the set $\mathcal{D}_C $ is open by definition. For $\sigma>0$, define 
\begin{align*}
    \mathcal{N}:= \{\text{bd}(\mathcal{D})\oplus\mathcal{B}_{\sigma}(0)\} \setminus \mathcal{D},
\end{align*}
where $\mathcal{N}$ is a ring of diameter $\sigma$ that surrounds the set $\mathcal{D}$ (golden ring in Figure \ref{fig:1}). We will use the set $\mathcal{N}$ to enforce that the value of the learned RHCBF $h(z)$ is negative on $\mathcal{N}$ to ensure that the set $\mathcal{C}$ is contained within the set $\mathcal{D}$. Hence, also assume that points $Z_N = \{z^i\}_{i=1}^{N_u}$ are artificially sampled from $\mathcal{N}$, i.e., $z^i\in \mathcal{N}$, by for instance gridding.  While the set $\mathcal{C}$ defined in \eqref{eq:set_C} considers all $z\in \mathbb{R}^{n_z}$ such that $h(z) \geq 0$, we modify this definition slightly by restricting the domain to the set $\mathcal{N} \cup \mathcal{D}$. This is a natural restriction since we are learning a RHCBF from data sampled over the domain $\mathcal{N} \cup \mathcal{D}$, and we therefore instead consider learning a  \emph{local} RHCBF $h(z)$  over $\mathcal{D}$ with respect to the set
\begin{align} \label{eq:local_C}
\mathcal{C}:=\{z\in\mathcal{N} \cup \mathcal{D}\, \big{|} \, h(z)\ge 0\}.
\end{align}

Let $\text{Lip}(h)$ be the local Lipschitz constant of $h(z)$ on $\mathcal{D}\cup\mathcal{N}$ and let us  define the functions
\begin{align*}
    q_c(z,u_c,t)&:=\langle\nabla h(z), \hat{W}_c(z,t,u_c)) \rangle -\|\nabla h(z)\|_\star\Delta_c(z,t,u_c)+\alpha(h(z))\\
    q_d(z,u_d,t)&:= h( \hat{W}_d(z,t,u_d))-\overline{\text{Lip}}\Delta_d(z,t,u_d)
\end{align*}
according to \eqref{eq:cbf_const_hybrid} and \eqref{eq:cbf_disc_hybrid}, but where $\overline{\text{Lip}}\in \mathbb{R}_{\ge 0}$ is a positive constant that we would like to upper bound $\text{Lip}(h,\hat{W}_d(z,t,u_d),\Delta_d(z,t,u_d))$. We further use the notation $\text{Lip}(q_c,z,\epsilon|u_c,t)$ and $\text{Lip}(q_d,z,\epsilon|u_c,t)$ to denote the local Lipschitz constants of the functions $q_c(z,u_c,t)$ and $q_d(z,u_d,t)$ for fixed $u_c$, $u_d$, and $t$, respectively, within the set $\mathcal{B}_{\epsilon}(z)$. Let us also denote by $\text{Bnd}(q_c|\mathcal{B}_{\epsilon}(z),u_c)$ and $\text{Bnd}(q_c|\mathcal{B}_{\epsilon}(z),u_d)$ the bounds on the difference between functions $q_c$ and $q_d$ for different $t$. In other words, for each $\bar{z}\in \mathcal{B}_{\epsilon}(z)$, it holds that
\begin{align*}
    |q_c(\bar{z},u_c,t')-q_c(\bar{z},u_c,t'')|\le \text{Bnd}(q_c|\mathcal{B}_{\epsilon}(z),u_c), \forall t',t''\ge 0,\\
    |q_d(\bar{z},u_d,t')-q_d(\bar{z},u_d,t'')|\le \text{Bnd}(q_d|\mathcal{B}_{\epsilon}(z),u_d), \forall t',t''\ge 0.
\end{align*}
The bounds $\text{Bnd}(q_c|\mathcal{B}_{\epsilon}(z),u_c)$ and $\text{Bnd}(q_d|\mathcal{B}_{\epsilon}(z),u_d)$ exist as the function $h$ is continuous and   the functions  $\hat{W}_c(z,t,u_c)$, $\hat{W}_d(z,t,u_d)$, $\Delta_c(z,t,u_c)$, and $\Delta_d(z,t,u_d)$ are assumed to be bounded in $t$. We later show that this is a natural assumption to obtain formal guarantees on our learned RHCBF $h(z)$ from finite data sets $Z_\text{dyn}^c$ and $Z_\text{dyn}^d$ since it is not possible to sample the time domain $\mathbb{R}_{\ge 0}$ densely with a finite number of samples. Naturally, these bounds can be neglected when \eqref{eq:hybrid_control_systems} does not depend on $t$.

We now propose an optimization problem for learning a  RHCBF, and then prove its correctness.  We solve
\begin{subequations}\label{eq:opt}
\begin{flalign}
    &\min_{h \in \mathcal{H}} \:\: \|h\|   \: \nonumber \\
    &\mathrm{s.t.}~~ h(z^i) \geq \gamma_\text{safe}, \: \forall z^i \in Z_\text{safe}^c\cup Z_\text{safe}^d \label{eq:cons_1}\\
    &~~~~~~ h(z^i) \leq -\gamma_\text{unsafe}, \: \forall z^i \in Z_N  \label{eq:cons_2}\\
    &~~~~~~ \text{Lip}(h,z^i,\bar{\epsilon}) \leq L_h, \: \forall z^i \in Z_N   \label{eq:lip1}\\
    &~~~~~~  q_c(z^i,u^i_c,t^i) \geq \gamma_{\text{dyn}}^c \label{eq:cons_3}\\
    &~~~~~~ \text{Bnd}(q_c|\mathcal{B}_{\epsilon_c}(z^i),u_c^i)\le M_q^c\label{eq:bnd2_c}\\
    &~~~~~~ \text{Lip}(q_c,z^i,\epsilon_c|u_c^i,t^i) \leq L_q^c, \: \forall (z^i, u^i_c,t^i) \in Z_\text{dyn}^c \label{eq:lip2_c}\\
     &~~~~~~  q_d(z^i,u^i_d,t^i) \geq \gamma_{\text{dyn}}^d  \label{eq:cons_4}\\
    &~~~~~~ \text{Bnd}(q_d|\mathcal{B}_{\epsilon_d}(z^i),u_d^i)\le M_q^d \label{eq:bnd2_d}\\
    &~~~~~~  \text{Lip}(h)\le \overline{\text{Lip}} \label{eq:lip2_dd}\\
    &~~~~~~ \text{Lip}(q_d,z^i,\epsilon_d|u_d^i,t^i) \leq L_q^d, \: \forall (z^i, u^i_d,t^i) \in Z_\text{dyn}^d \label{eq:lip2_d}
\end{flalign}
\end{subequations}
where $\mathcal{H}$ is a normed function space of at least twice continuously differentiable  functions and where the positive constants $\gamma_\text{safe}$, $\gamma_\text{unsafe}$, $\gamma_{\text{dyn}}^c$, $\gamma_{\text{dyn}}^d$, $L_h$, $L_q^c$, $L_q^d$, $\overline{\text{Lip}}$, $M_q^c$, and $M_q^d$ are
\emph{hyperparameters} determined by the data-sets $Z_\text{safe}^c$, $Z_\text{safe}^d$, and $Z_N$, which must be sufficiently dense (conditions given below).  We remark here that all hyperparameters can be positive functions of $z^i$ instead of being global constants to obtain  less conservative conditions. In brief, equation \eqref{eq:cons_1} enforces $h$ to be positive with margin $\gamma_\text{safe}$ within the safe regions $Z_\text{safe}^c$ and $Z_\text{safe}^d$, while equation \eqref{eq:cons_2} enforces $h$ to be negative with margin  $-\gamma_\text{unsafe}$ within the as unsafe labelled region $Z_N$. Equations \eqref{eq:cons_3} and \eqref{eq:cons_4} together with \eqref{eq:lip2_dd} enforce the derivative and jump conditions \eqref{eq:cbf_const_hybrid} and \eqref{eq:cbf_disc_hybrid} with margins $\gamma_{\text{dyn}}^c$ and $\gamma_{\text{dyn}}^d$, respectively. The Lipschitz and boundedness constraints \eqref{eq:lip1}, \eqref{eq:bnd2_c}, \eqref{eq:lip2_c}, \eqref{eq:bnd2_d}, and \eqref{eq:lip2_d} are used in the remainder to generalize beyond data-points. 

For general function classes $\mathcal{H} := \{h(z; \theta) | \theta\in\Theta\}$ where $\Theta$ is a set of parameters, the optimization problem posed in \eqref{eq:opt} is nonconvex and challenging to solve.  Inspired by \cite{chamon2020probably}, we leverage recent results in constrained PAC learning towards formulating an algorithm that can be used to efficiently solve a relaxed version of \eqref{eq:opt}.   To this end, we first define the empirical Lagrangian for the optimization problem in \eqref{eq:opt} with respect to the parameterized function class $\mathcal{H}$ as follows:
 \begin{align*}
     &L(\theta, \lambda_\text{safe}, \lambda_\text{unsafe}, \lambda_{\text{dyn},c}, \lambda_{\text{dyn},d}) := \|h\| + \frac{1}{|Z_{\text{safe}}^c \cup Z_{\text{safe}}^d|} \hspace{5pt} \cdot \hspace{5pt} \sum_{\mathclap{z_i\in Z_{\text{safe}}^c \cup Z_{\text{safe}}^d}} \lambda_{\text{safe}}^i \left[ \gamma_\text{safe} - h(z^i) \right]_+ \notag \\
     &\qquad + \frac{1}{|Z_N|} \sum_{z^i\in Z_N} \lambda_{\text{unsafe}}^i \left[ h(z^i) + \gamma_\text{unsafe} \right]_+  + \frac{1}{|Z_\text{dyn}^c|} \hspace{5pt} \cdot \hspace{5pt} \sum_{\mathclap{(z^i, u^i, t^i) \in Z_\text{dyn}^c}} \lambda_{\text{dyn},c}^i \left[\gamma_\text{dyn}^c - q_c(z^i, u^i, t^i) \right]_+ \notag \\
     &\qquad + \frac{1}{|Z_\text{dyn}^d|} \hspace{5pt} \cdot \hspace{5pt} \sum_{\mathclap{(z^i, u_d^i, t^i) \in Z_\text{dyn}^d}} \lambda_{\text{dyn},d}^i \left[ \gamma_\text{dyn}^d - q_d(z^i, u_d^i, t^i) \right]_+
 \end{align*}
Here $\lambda := (\lambda_\text{safe}, \lambda_\text{unsafe}, \lambda_{\text{dyn},c}, \lambda_{\text{dyn},d})$ denotes the collection of dual variables corresponding to the constraints of \eqref{eq:opt}, and $[ r ]_+ := \max\{0, r\}$.  Note that we have implicitly relaxed the primal problem \eqref{eq:opt} by removing the Lipschitz and boundedness constraints \eqref{eq:lip1}, \eqref{eq:bnd2_c}, \eqref{eq:lip2_c}, \eqref{eq:bnd2_d}, \eqref{eq:lip2_dd}, and \eqref{eq:lip2_d} in this Lagrangian.  In general, we rely on a post-hoc validation scheme to ensure that these constraints are satisfied. A further discussion of this scheme for the Lipschitz constraints is provided in \cite{lindemann2020learning}, while a validation scheme for the boundedness constraints can be derived similarly.

Next, we use this Lagrangian to formulate the empirical dual of the relaxed version of \eqref{eq:opt}:
\begin{align}
    \max_{\lambda \geq 0} \:  \min_\theta L(\theta, \lambda_\text{safe}, \lambda_\text{unsafe}, \lambda_{\text{dyn},c}, \lambda_{\text{dyn},d}). \label{eq:dual}
\end{align}
Due to the nonconvexity of the primal problem for a wide variety of function classes $\mathcal{H}$ commonly used in deep learning (e.g.\ DNNs), strong duality does not hold in general, and thus a saddle point of \eqref{eq:dual} does not correspond to a solution for our relaxation of \eqref{eq:opt}.  We therefore propose Algorithm \ref{alg:alg} inspired by \cite{chamon2020probably} where $\eta,\beta>0$ are gradient step sizes, which were chosen by grid search.  Further, we follow \cite{chamon2020probably} by initializing each dual variable with $\boldsymbol{1}$ (i.e.\ the all-ones vector).

\begin{algorithm}
\centering  
\begin{algorithmic}[1]
    \For{Epoch $e=1, \dots, E$}
        \State Solve one step of inner minimization \begin{equation}
            \theta \gets \theta - \eta \nabla_\theta L(\theta, \lambda) \notag
        \end{equation}
        \State Solve one step of outer maximization 
        \begin{gather*}
            \lambda_\text{safe}^i \gets \left[ \lambda_\text{safe}^i + \beta \left(\gamma_\text{safe} - h(z^i) \right) \right]_+ \notag \\
            \lambda_\text{unsafe}^i \gets \left[ \lambda_\text{unsafe}^i + \beta\left(h(z^i) + \gamma_\text{unsafe} \right) \right]_+ \\
            \lambda_{\text{dyn},c}^i \gets \left[ \lambda_{\text{dyn},c}^i + \beta\left(\gamma_\text{dyn}^c - q_c(z^i, u^i, t^i) \right) \right]_+ \\
            \lambda_{\text{dyn},d}^i \gets \left[ \lambda_{\text{dyn},d}^i + \beta \left(\gamma_\text{dyn}^d - q_d(z^i, u_d^i, t^i)\right)\right]_+
        \end{gather*}
    \EndFor
    \State \Return $\theta$
\end{algorithmic}
\caption{Primal Dual Iteration}
\label{alg:alg}
\end{algorithm}

\subsection{Formal Correctness Guarantees} 
We show correctness of the learned HCBF $h(z)$  obtained from \eqref{eq:opt} in two steps by: 1) showing that the certified safe set \eqref{eq:local_C} is contained within the geometric safe set, i.e., that $\mathcal{C}\subset \mathcal{D}\subseteq \mathcal{S}$, and 2) proving that $h(z)$ is a local RHCBF by ensuring that the set $\mathcal{C}$ is robustly controlled forward invariant. As remarked, we assume in the remainder that $L_h$, $L_q^c$, $L_q^d$, $M_q^c$, and $M_q^d$ are functions of $z$.

\subsubsection{1) Guaranteeing $\mathcal{C}\subset \mathcal{D}\subseteq \mathcal{S}$: } We say that $Z_N $ is an $\bar{\epsilon}$-net of $\mathcal{N}$ if for all $z \in \mathcal{N}$,  there exists $z^i\in Z_N$ such that $\|z^i-z\|\leq \bar{\epsilon}$. The following result is directly taken from \cite{lindemann2020learning}.
\begin{proposition}\label{prop:1}
Let $h(z)$ be locally Lipschitz continuous and satisfy the constraints \eqref{eq:cons_2} and \eqref{eq:lip1}. Let $\gamma_\text{unsafe}>0$ and $Z_N $ be an $\bar{\epsilon}$-net of $\mathcal{N}$ with $\bar{\epsilon}<\gamma_\text{unsafe}/L_h(z^i)$ for all $z^i \in Z_N$. Then we have that $h(z)<0$ for all $z\in\mathcal{N}$.
\end{proposition}

Note that the constraint \eqref{eq:cons_1} pushes $h(z^i)>0$ for all safe data-point $z^i \in Z_\text{safe}^c\cup Z_\text{safe}^d$ so that the set $\mathcal{C}$ is not empty. When the conditions in Proposition \ref{prop:1} hold, it then follows that $\mathcal{C}$ is inside $\mathcal{D}$ and hence inside of $\mathcal{S}$, i.e., $\mathcal{C}\subset \mathcal{D}\subseteq \mathcal{S}$, by the construction of $\mathcal{N}$. 

To avoid a disconnected set $\mathcal{C}$ that has holes, which would ultimately degrade control performance, we recall the following proposition from \cite{lindemann2020learning}. Note here that, by definition of $\mathcal{D}_C$ and $\mathcal{D}_D$, the sets $Z_\text{safe}^c$ and $Z_\text{safe}^d$ are ${\epsilon}_c$- and ${\epsilon}_d$-nets of $\mathcal{D}_C$ and $\mathcal{D}_D$, respectively.
\begin{proposition}\label{cor:1}
Let $h(z)$ be locally Lipschitz continuous and satisfy the constraint \eqref{eq:cons_1}. Let $\gamma_\text{safe}>0$ with $\epsilon_c\leq\gamma_\text{safe}/L_h(z^i)$ for all  $\ z^i \in Z_\text{safe}^c$ and $\epsilon_d\leq\gamma_\text{safe}/L_h(z^i)$ for all  $\ z^i \in Z_\text{safe}^d$. Then we have that $h(z)\geq 0$ for all $z\in\mathcal{D}$.
\end{proposition}

The constraints \eqref{eq:cons_1} and \eqref{eq:cons_2} may lead to infeasibility of \eqref{eq:opt}. As in \cite{lindemann2020learning}, one may instead enforce constraint \eqref{eq:cons_1} on smaller sets $\bar{Z}_\text{safe}^c$ and $\bar{Z}_\text{safe}^d$ with $\bar{Z}_\text{safe}^c\subset Z_\text{safe}^c$ and $\bar{Z}_\text{safe}^d\subset Z_\text{safe}^d$ by removing points $z^i$ close to the set $\mathcal{N}$ to allow for smoother RHCBFs to be learned at the expense of a smaller invariant safe set $\mathcal{C}$.

\subsubsection{2) Guaranteeing local RHCBF: }  We next provide conditions guaranteeing that the learned RHCBF satisfies the flow constraint \eqref{eq:cbf_const_hybrid} for all $(z,t) \in \mathcal{D}_C\times\mathbb{R}_{\ge 0}$ and the jump constraint \eqref{eq:cbf_disc_hybrid} for all $(z,t) \in \mathcal{D}_D\times\mathbb{R}_{\ge 0}$.

\begin{proposition}\label{prop:2}
Let $h(z)$ be locally Lipschitz continuous in $z$ and let $q_c(z,u_c,t)$ and $q_d(z,u_d,t)$ satisfy the Lipschitz constraints \eqref{eq:lip2_c} and \eqref{eq:lip2_d} as well as the constraints \eqref{eq:cons_3} and \eqref{eq:cons_4}. Let also the boundedness constraints \eqref{eq:bnd2_c} and \eqref{eq:bnd2_d} hold. Let $\gamma_\text{dyn}^c,\gamma_\text{dyn}^d>0$, and assume that 
\begin{itemize}
    \item  $\epsilon_c\leq (\gamma_\text{dyn}^c-M_q^c(z^i))/L_q^c(z^i)$ for all $(z^i,u_c^i,t^i)\in Z_\text{dyn}^c$, 
    \item  $\epsilon_d\leq (\gamma_\text{dyn}^d-M_q^d(z^i))/L_q^d(z^i)$ for all $(z^i,u_d^i,t^i)\in Z_\text{dyn}^d$.
\end{itemize}   
Then, for each $(z,t) \in \mathcal{D}_C\times \mathbb{R}_{\ge 0}$, there exists a $u_c\in\mathcal{U}_c$ such that $q_c(z,u_c,t)\geq 0$, and, for each $(z,t) \in \mathcal{D}_D\times \mathbb{R}_{\ge 0}$, there exists a $u_d\in\mathcal{U}_d$ such that $q_d(z,u_d,t)\geq 0$.
\begin{proof}
 Note first that, for each $z\in\mathcal{D}_C$, it follows that there exists a pair $(z^i,u_c^i,t^i) \in Z_\text{dyn}^c$ satisfying $\|z-z^i\|\leq \epsilon_c$ since $Z_\text{safe}^c$ is an $\epsilon_c$-net of $\mathcal{D}_C$. For any pair $(z,t)\in\mathcal{D}_C\times\mathbb{R}_{\ge 0}$, we can now select such a pair $(z^i,u_c^i,t^i) \in Z_\text{dyn}^c$ satisfying $\|z-z^i\|\leq \epsilon_c$ for which it follows that
\begin{align*}
    0&\overset{(a)}{\leq} q_c(z^i,u_c^i,t^i)-\gamma_\text{dyn}^c\\
    &\leq|q_c(z^i,u_c^i,t^i)-q_c(z,u_c^i,t^i)|+q_c(z,u_c^i,t^i)-\gamma_\text{dyn}^c\\
    &\overset{(b)}{\leq}L_q^c(z^i)\|z^i-z\|+q_c(z,u_c^i,t^i)-\gamma_\text{dyn}^c\\
    &\overset{(c)}{\leq}L_q^c(z^i)\epsilon_c+q_c(z,u_c^i,t^i)-\gamma_\text{dyn}^c\\
    &\le L_q^c(z^i)\epsilon_c+|q_c(z,u_c^i,t^i)-q_c(z,u_c^i,t)|+q_c(z,u_c^i,t)-\gamma_\text{dyn}^c\\
    &\overset{(d)}{\leq}L_q^c(z^i)\epsilon_c+M_q^c(z^i)+q_c(z,u_c^i,t)-\gamma_\text{dyn}^c\overset{(e)}{\leq}q_c(z,u_c^i,t).
\end{align*}
Inequality $(a)$ follows from the constraint \eqref{eq:cons_3}. Inequality $(b)$ follows by the upper bound $L_q^c(z^i)$ on the  Lipschitz constant $\text{Lip}(q_c,z^i,\epsilon_c|u_c^i,t^i)$ of $q_c(z,u_c^i,t^i)$ within the set $\mathcal{B}_{\epsilon_c}(z^i)$ from constraint \eqref{eq:lip2_c}. Inequality $(c)$ follows again since $Z_\text{safe}^c$ is an $\epsilon_c$-net of $\mathcal{D}_C$. Inequality $(d)$ follows from constraint \eqref{eq:bnd2_c}. Inequality $(e)$ follows simply by the assumption that $\epsilon_c\leq(\gamma_\text{dyn}^c-M_q^c(z^i))/L_q^c(z^i)$ for all $z^i \in Z_\text{safe}^c$. Consequently, $q_c(z,u_c^i,t)\ge 0$ for all $(z,t)\in\mathcal{D}_C\times\mathbb{R}_{\ge 0}$. The same analysis holds for all $(z,t)\in \mathcal{D}_D\times\mathbb{R}_{\ge 0}$, so that $q_d(z,u_d^i,t)\ge 0$ for all $(z,t)\in\mathcal{D}_D\times\mathbb{R}_{\ge 0}$.
 \end{proof}
\end{proposition}

From Proposition \eqref{prop:2}, it can now be concluded that the flow constraint \eqref{eq:cbf_const_hybrid} is always feasible. Additionally, Proposition \eqref{prop:2} in conjunction with the constraint  \eqref{eq:lip2_dd} ensures that the jump constraint \eqref{eq:cbf_disc_hybrid} is always feasible.

\section{Simulations} \label{sec:case_studies}

\begin{figure*}
    \centering
    \includegraphics[width=\textwidth]{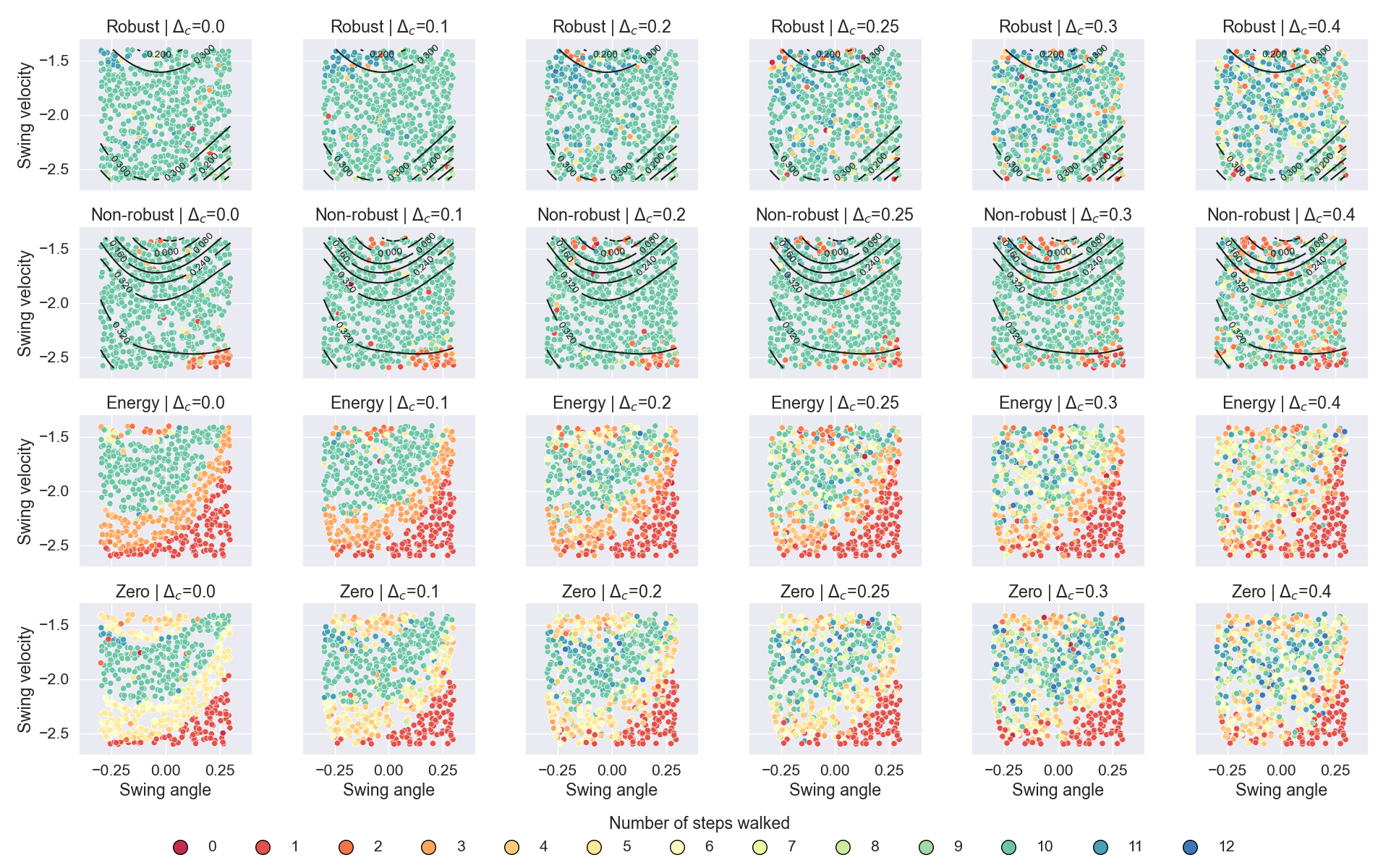}
    \caption{Each dot corresponds to an initial condition of the compass gait walker where the swing leg is varied, while the stance leg is fixed. The color of each dot corresponds to the number of steps under a particular control law and when the system dynamics are subject to additive noise with upper bound $\Delta_c$.  The columns correspond to different test-time values of $\Delta_c$, and the rows correspond to different control policies (`Robust' for RHCBF, 'Non-robust' for the HCBF from \cite{lindemann2020learning}, 'Energy' for the energy-based expert controller of \cite{goswami1996limit}, and 'Zero' for no actuation). The black lines in the first two rows show the levels sets of the HCBF and RHCBF.}  
    \label{fig:compass-gait-ics}
\end{figure*}

To demonstrate the utility of learning RHCBFs from expert demonstrations, we consider the compass gait walker dynamical system \cite{goswami1996limit}, which describes a passive bipedal robot walking down an inclined plane at a constant velocity.  When written in control-affine form, this systems is characterized by a four-dimensional state $z := [\theta_{\text{stance}}, \theta_{\text{swing}}, \dot{\theta}_{\text{stance}}, \dot{\theta}_{\text{swing}}]$ consisting of the angle and angular velocity of each leg.  In this notation, the ``stance'' foot corresponds to the foot that is in contact with the ground as the compass gait walker makes its descent down the ramp; hence, the ``swing'' foot refers to the foot that is not in contact with the ramp at a particular instant in time. To simulate this system, we numerically
integrate the hybrid dynamics using an integrator implementation inspired by \cite{drake}.   Our code is available at \href{https://github.com/unstable-zeros/learning-hcbfs}{https://github.com/unstable-zeros/learning-hcbfs}. In our simulations, the walker's initial stance and swing legs are its respective left and right leg.  To improve the walking capabilities, we add actuation to hip and ankle joints of the stance leg.  To collect safe expert data, we use the energy-based controller of~\cite{goswami1997limit}.   

Learning and analyzing a RHCBF for the compass gait walker poses several challenges, including the well-known sensitivity of the compass gait walker to its initial conditions (see last row in Figure \ref{fig:compass-gait-ics} when no actuation is applied).  To facilitate a meaningful visualization of the four-dimensional state space, when collecting expert data, we fix the stance leg initial condition to the point $[\theta_{\text{stance}}, \dot{\theta}_{\text{stance}}] = [0, 0.4]$ on the passive limit cycle, and vary the initial condition of the swing leg by adding uniform noise to corresponding passive limit cycle state $[\theta_{\text{swing}}, \dot{\theta}_{\text{swing}}] = [0, -2.0]$ of the swing leg. 

\begin{figure}
    \centering
    \includegraphics[scale=0.5]{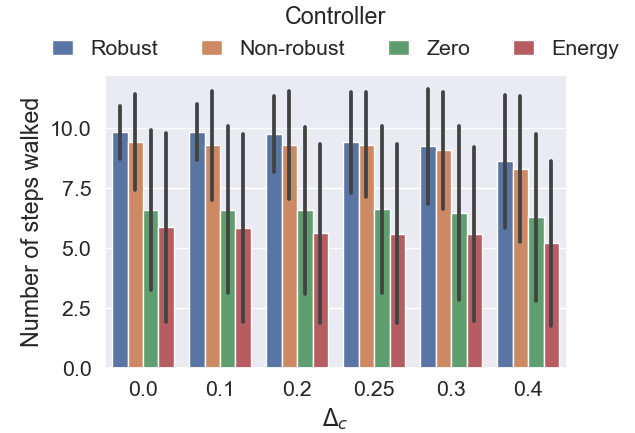}
    \caption{Average number of steps of the compass gait walker. }
    \label{fig:my_label2}
\end{figure}

We first perturb the system by uniform noise and collect expert data with $\Delta_c(z,t,u_c):=0.25$. We then train a HCBF based on \cite{lindemann2020learning} and a RHCBF as proposed in this paper with $\Delta_c(z,t,u_c):=0.25$. We parameterize both the candidate HCBF and RHCBF with a two-hidden-layer fully-connected neural network with tanh activations and 32 (resp.\ 16) neurons in the first (resp. and second) hidden layers.
For the primal-dual iteration, we use parameters $E = 30000$, $\eta = 0.005$, and $\beta = 0.05$. For noise where $\Delta_c(z,t,u_c)$ ranges between $0$ and $0.4$, the results are shown in the second and first row of Figure \ref{fig:compass-gait-ics}, respectively. The energy-based expert controller is shown in the third row for comparison. Figure \ref{fig:my_label2} summarizes the average number of steps. Note that the HCBF controller is already much more robust than the energy-based controller due to the use of the margins $\gamma_{\text{dyn}}^c$ and $\gamma_{\text{dyn}}^d$, but that our RHCBF controller further improves the controllers ability to reject noise.

\begin{figure*}
    \centering
    \includegraphics[width=\textwidth]{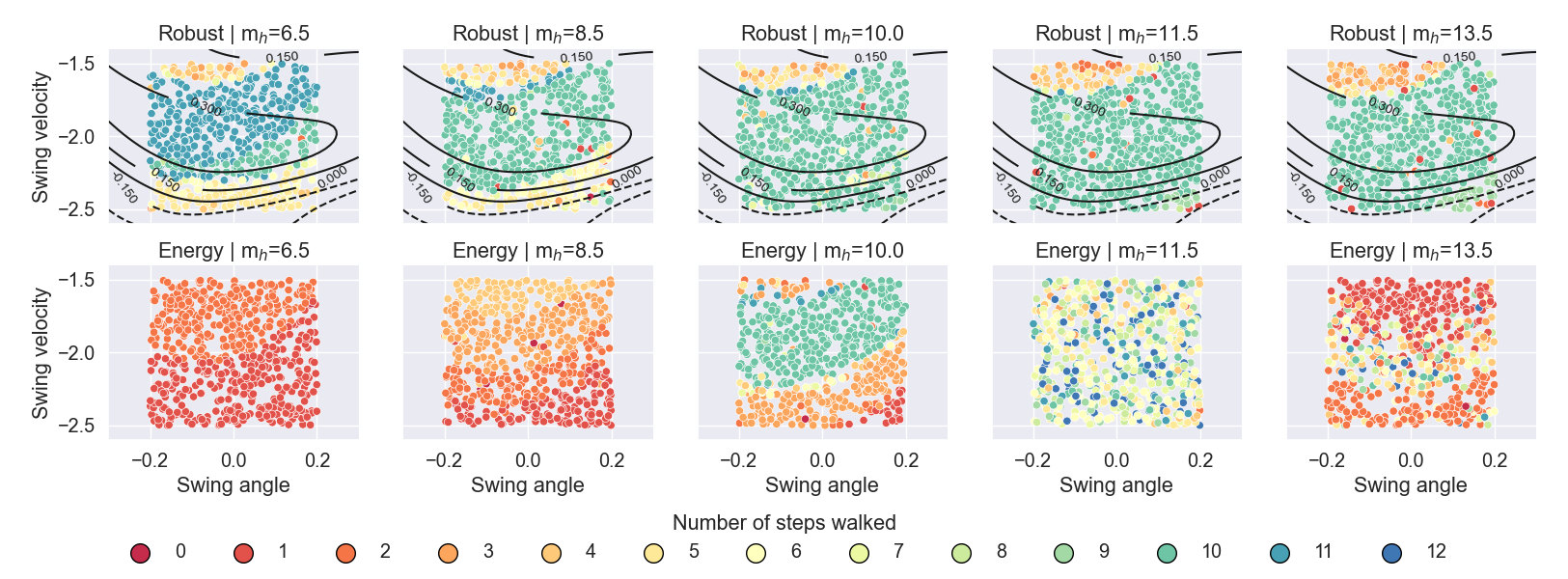}
    \caption{Simulation results for the compass gait walker when the hip mass is uncertain. Here `Robust' stands again for the RHCBF that is used with parameters $\hat{W}(z,u_c)=f(z;10)+g(z;10)u_c$ and $\Delta_c(z,u_c)=0.1$, while 'Energy' stands for the energy-based expert controller of \cite{goswami1996limit} that is used with the nominal dynamics $f(z;10)+g(z;10)u_c$. }
    \label{fig:my_label3}
\end{figure*}

We next consider the hip mass of the compass gait walker to be uncertain (this corresponds to the parameter $m_H$ in \cite{goswami1996limit}). We have used the energy-based expert controller to collect safe data for $m_H$ varying between $9.25$ and $10.75$. During training of the RHCBF, we have set $\hat{W}(z,u_c)=f(z;10)+g(z;10)u_c$ and $\Delta_c(z,u_c)=0.1$ where $f(z;m_H)$ and $g(z;m_H)$ are the internal and input dynamics of the compass gait walker during flows for the particular hip mass of $m_H$. We remark that $\Delta_c(z,u_c)$ can in principle be determined more accurately by estimating $\sup_{m_H\in[9.25,10.75]}\|f(z;m_H)-f(z;10)+g(z;m_H)-g(z;10)\|u_c$. We again parameterize  the candidate RHCBF with a two-hidden-layer fully-connected neural network with tanh activations and 32 (resp.\ 16) neurons in the first (resp. and second) hidden layers. For the primal-dual iteration, we again used parameters $E = 30000$, $\eta = 0.005$, and $\beta = 0.05$. The  obtained results are shown in Figure \ref{fig:my_label3}.

\begin{figure}
    \centering
    \includegraphics[scale=0.5]{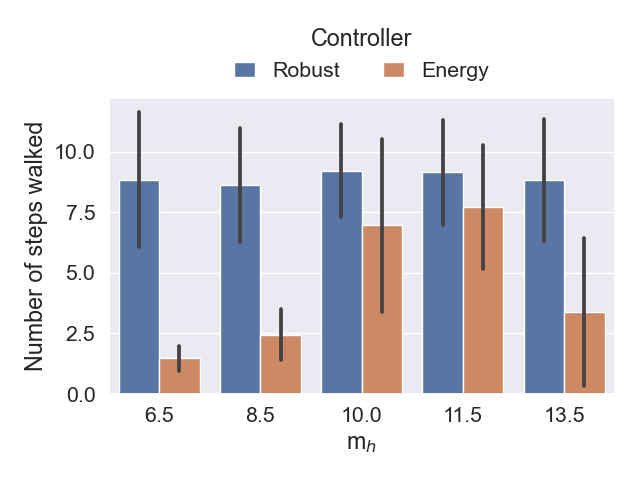}
    \caption{Average number of steps walked by the compass gait walker given an uncertain hip mass $m_h$.}
    \label{fig:hip-mass-num-steps}
\end{figure}

\section{Conclusion}
\label{sec:conclusion}
This paper first proposed robust hybrid control barrier functions  to synthesize control laws that ensure robust safety. We then formulated an optimization problem to learn such functions from data along with sufficient conditions on the data that ensure correctness of our approach.

\clearpage

\newpage
\bibliographystyle{IEEEtran}
\bibliography{literature}

\newpage

\end{document}